\documentclass[a4paper,11pt]{article}
\usepackage{jheppub} 
\usepackage[utf8]{inputenc}
\usepackage{graphicx}
\usepackage{subfigure}
\usepackage{setspace}
\usepackage{amsmath}
\usepackage{commath}
\usepackage{amsthm}
\usepackage{color}
\usepackage{braket}
\usepackage{color,soul}
\usepackage{appendix}
\usepackage{amssymb}
\allowdisplaybreaks
\usepackage[dvipsnames]{xcolor}

\usepackage{lineno}

\usepackage{mathtools}
\title{Thermalization of isolated quantum many-body system  and the role of entanglement}

\author[a, b]{Tanmay Saha,}
\author[c, d]{Pratik Ghosal,}
\author[d]{Pratapaditya Bej,}
\author[d]{Abhishek Banerjee,}
\author[e, d]{Prasenjit Deb}
\affiliation[a]{Optics \& Quantum Information Group, The Institute of Mathematical Sciences, CIT Campus, Taramani, Chennai 600113, India}
\affiliation[b]{Homi Bhabha National Institute, Training School Complex, Anushakti Nagar, Mumbai 400085, India}
\affiliation[c]{Department of Physics of Complex Systems, S. N. Bose National Centre for Basic Sciences, Block JD, Sector III, Salt Lake, Kolkata 700106, India}
\affiliation[d]{Department of Physical Sciences, Bose Institute, EN 80, Sector V, Bidhannagar, Kolkata 700091, India}
\affiliation[e]{Center for Quantum Engineering, Research, and Education, TCG CREST, Block EM, Bidhan Nagar, Kolkata - 700091, India.}
\emailAdd{sahatanmay@imsc.res.in}
\emailAdd{ghoshal.pratik00@gmail.com}
\emailAdd{pratap6906@gmail.com}
\emailAdd{abhishekbanerjee2001@gmail.com}
\emailAdd{devprasen@gmail.com}

\abstract{Thermalization of an isolated quantum system has been a nontrivial problem since the early days of quantum mechanics. In generic isolated quantum systems, nonequilibrium dynamics is expected to result in thermalization, indicating the emergence of statistical mechanics from quantum dynamics. However, what feature of a many-body quantum system facilitates quantum thermalization is still not well understood. Recent experimental advancements have shown that entanglement may act as a thermalizing agent, not universally but particularly. Here, we theoretically show that the thermal averages of an observable in an isolated many-body quantum system with a large number of degrees of freedom emerge from the entangled energy eigenstates of the system. In particular, we show that the expectation values of an observable in entangled energy eigenstates and its marginals are equivalent to the microcanonical and canonical averages of the observable.}

\begin{document}
\maketitle
\flushbottom

\section{Introduction}
\label{intro}
A prerequisite for statistical mechanics is the maximization of entropy in a system at thermal equilibrium. In other words, when a system gets thermalized one can find out the values of corresponding physical observables and thermodynamic functions from its statistical description or representative ensemble. However, an isolated quantum many-body system initialized in a pure state remains pure during unitary evolution, and in this sense, it has zero entropy. Then, what is the mechanism through which such a quantum system, whose initial state is pure, gets thermalized and quantum statistical mechanics emerges from it? Thermalization of an isolated quantum system and the emergence of statistical ensembles from its unitary time evolution has been a fascinating problem since the early days of quantum  mechanics\cite{von, PhysRevLett.54.1879,PhysRevLett.108.110601,PhysRevLett.115.100402,tasaki2016typicality,Gogolin_2016,BORGONOVI20161,Mori_2018}.
In the classical scenario, the assumption of \textit{ergodicity} leads to statistical mechanics\cite{ma1985statistical}. However, the notion of \textit{ergodicity} adopted for classical systems has failed in leading to a similar conclusion in the quantum regime despite numerous attempts\cite{von,farquhar1964ergodic,fierz}. Most of the works have emphasized the need for coupling with an external heat bath\cite{PhysRev.107.333}, which is being done traditionally, in order to obtain statistical mechanics. Later, it has been shown in the seminal works by Deutsch\cite{PhysRevA.43.2046} and Srednicki\cite{PhysRevE.50.888} that a finite but very small perturbation in the form of a random matrix may lead to such a temporal evolution of the system that the time averages of observables are in agreement with the microcanonical ensemble, commonly known as the \textit{eigenstate thermalization hypothesis} (ETH). The name itself signifies that thermalization happens at the level of individual eigenstates.

In the last decade, experimental developments\cite{kinoshita2006quantum,hofferberth2007non,trotzky2012probing,cheneau2012light,langen2013local,islam2015measuring,langen1} have made precise simulation of unitary evolution of quantum many-body systems and important experimental studies of thermalization possible, stimulating theoretical interest. Discussion about those theoretical works is beyond the scope of this paper, however, one can find those in \cite{PhysRevLett.108.110601,rigol2008thermalization,PhysRevLett.125.070605,PhysRevLett.98.050405} and the references therein. A generic isolated quantum many-body system thermalizes to a microcanonical distribution consistent with their energy density\cite{rigol2008thermalization}, and the experimental results are consistent with this fact. The mechanism behind this is eigenstate thermalization, as prescribed by the \textit{eigenstate thermalization hypothesis}. Though ETH successfully describes the thermalization of a generic isolated system, integrable systems possessing extensive sets of non-trivial conserved quantities do not follow it. As a result, in general, integrable systems do not thermalize\cite{PhysRevLett.125.070605}, rather they do equilibrate. To describe such intergrable systems after equilibration generalized Gibbs ensembles (GGEs) are used\cite{PhysRevLett.98.050405}. In the last few years, a lot of research has been carried out to understand the thermalization of both integrable and non-integrable systems. However, what feature of many-body quantum system helps in quantum thermalization is not clear yet. Recently, experimental studies\cite{doi:10.1126/science.aaf6725} with ultra-cold atoms have confirmed that entanglement\cite{schrodinger1935gegenwartige,PhysRevA.40.4277,RevModPhys.81.865} may act as a thermalizing agent in isolated quantum many-body system. The confirmation comes from the simultaneous measurement of entanglement entropy\cite{von1955mathematical,nielsen} and thermal averages of observables of the subsystems. As the system's state, initialized in a pure one, moves towards thermal equilibrium, its entanglement entropy starts to grow. The growth of entanglement entropy with respect to time and size of the subsystems has been studied in \cite{doi:10.1126/science.aaf6725}. Later, using  standard quasiparticle picture the entanglement dynamics in the space-time scaling limit has been studied\cite{doi:10.1073/pnas.1703516114}. However, the mere presence of entanglement does not guarantee the thermalization of arbitrary isolated many-body quantum system. For example, in the case of the integrable XXZ spin chain model, ETH does not apply but entanglement develops from time evolution. Another example includes systems exhibiting many-body localization (MBL) where thermalization does not occur but each eigenstate displays area-law entanglement\cite{RevModPhys.91.021001}.

From the existing studies, it can be concluded that entanglement is not sufficient for the thermalization of an isolated many-body quantum system, what type of system is considered also matters. Similarly, given an isolated many-body quantum system, any arbitrary entanglement may not guarantee the thermalization of the system. Here we study how entanglement leads to thermalization in an isolated many-body thermalizable system. We show that the expectation values of an observable in the energy eigenstates of the global system and local subsystems are equivalent to the thermal averages of the observable due to the presence of entanglement.\\

The rest of the article is arranged as follows: In Section \ref{sec2}, we briefly review the eigenstate thermalization of a generic isolated many-body quantum system. Section \ref{res} demonstrates our findings. Finally, in Section \ref{con} we conclude our work.

\section{Eigenstate thermalization of a generic isolated quantum system}\label{sec2}
First introduced by Deutsch\cite{PhysRevA.43.2046} and coined by Srednicki\cite{PhysRevE.50.888}, ETH aims to recover the results of quantum statistical mechanics from a closed quantum system. For this purpose, a small perturbation Hamiltonian $\hat{H}_{int}$, in the form of a random matrix, is added to the system, and the system is allowed to evolve under the total Hamiltonian $\hat{H}=\hat{H_0}+\hat{H}_{int}$. Here, $\hat{H_0}$ is the initial Hamiltonian of the system. At first glance, it seems that due to the external perturbation, the system having a well-defined energy may cause macroscopic fluctuation in its energy. However, it can be easily shown that after the perturbation has been turned off the ratio $\Delta E/E$ decreases as $1/\sqrt D$, where $\Delta E$ is the spread of the total energy $E$ and $D$ is the degrees of freedom. Therefore, the eigenstate occupation probabilities remain localized around a small range of energies. The elements of $\hat{H}_{int}$ are chosen randomly from a real symmetric Gaussian ensemble, and in the basis of the eigenvectors of $\hat{H_0}$ these are represented as,
\begin{eqnarray}\label{sec_two_1}
h_{ij}\equiv \langle E_i\lvert \hat{H}_{int}\lvert E_j\rangle,~~
\langle h_{ij}h_{kl}\rangle=\epsilon^2\delta_{ik}\delta_{jl}.
\end{eqnarray}
Such modeling of the perturbation Hamiltonian yields that in the limit of large $D$, small but finite $\epsilon$ will couple many neighboring levels within the range $E$ and $E\pm \Delta E$. This coupling between energy levels occurs because, at fixed energy per particle, the separation between levels decreases exponentially with $D$ and becomes arbitrarily small. The number of energy levels in the window $\Delta E$
is proportional to $\Delta E ~\mbox{exp}[S(E)]$\cite{landau2013statistical}, $S(E)$ being the total entropy at total energy $E$, and there is a large range of values for $\epsilon$ which will couple these large number of levels. It is expected that for large $D$, the coupling energy $\epsilon$ can be made much smaller than the energy per particle, and have a large effect on eigenvectors. The eigenvectors of $\hat{H}$ should then coherently mix the eigenvectors of $\hat{H_0}$ within a window $\Delta E$. This mixing of the unperturbed eigenvectors finally gives rise to ergodicity.

Let, $\lvert \psi_i\rangle$ be an eigenvector of the total Hamiltonian $\hat{H}$, $\{\lvert \phi_j\rangle , j=1~ \mbox{to}~ D\}$ be the eigenvectors of the unperturbed Hamiltonian $\hat{H_0}$ and an initial state of the system be,
\begin{equation}\label{sec_two_2}
\lvert \Psi\rangle=\sum^D_{i=1}~C_i\lvert \psi_i\rangle,
\end{equation}
where, $\sum_i~ \lvert C_i\lvert^2=1$.
Then, ETH\cite{PhysRevA.43.2046} implies (for the detailed analysis and updated review see\cite{deutsch_unpublished,Deutsch_2010,Deutsch_2018}):
\begin{equation}\label{sec_two_3}
\langle\langle \psi_i\lvert \hat{A} \lvert\psi_i\rangle\rangle_{rand}=\langle \hat{A}\rangle_{micro}(E_{i}),
\end{equation}
where, $E_{i}$ is the energy eigenvalue concomitant with the eigenvstate $\lvert \psi_i\rangle$ of $\hat{H}$.

In Eq.~(\ref{sec_two_3}), the averaging of the expectation value of an observable $\hat{A}$, a self-adjoint operator, is performed over different realizations of $\hat{H}_{int}$ [denoted by $\langle\cdot\cdot\cdot\rangle_{rand}$]. From this equation, it is clear that thermalization happens at the level of individual eigenstates. The variance $\Delta \hat{A}^2$, which can be written as $\Delta \hat{A}^2\equiv \langle\langle \psi_i\lvert \hat{A} \lvert\psi_i\rangle^2\rangle_{rand}-\langle\langle \psi_i\lvert \hat{A} \lvert\psi_i\rangle\rangle_{rand}^2$, decreases exponentially with $D$. Therefore, in the limit of large $D$, the infinite time average [denoted by $\langle\cdot\cdot\cdot\rangle_{t}$] of the expectation value of $\hat{A}$ is
\begin{eqnarray}\label{sec_two_4}
\langle\langle\langle \Psi(t)\lvert \hat{A} \lvert\Psi(t)\rangle\rangle_{t}\rangle_{rand} =\sum_i~ \lvert C_i\lvert^2\langle \hat{A}\rangle_{micro}(E_{i}). 
\end{eqnarray}

Given the mechanism of thermalization of an isolated quantum many-body system, one question naturally arises: what should be the property or properties of the system due to which thermalization will take place following that mechanism? In\cite{PhysRevE.50.888}, it has been shown that a closed quantum many-body system will thermalize according to eigenstate thermalization if \textit{Berry's conjecture} \cite{MVBerry_1977} holds for that system. On the other hand, \textit{Berry's conjecture} is found to be valid for eigenstates of sufficiently high energy in classical chaotic systems. Therefore, the implication of the validity of this conjecture for an isolated quantum system is that it has to be chaotic. 

Let, the state of such a chaotic quantum system at any instant of time $t$ is
\begin{equation}\label{sec_two_5}
\lvert \Psi(t)\rangle=\sum^D_{i=1}~C_i\mbox{e}^{-i E_i t}\lvert \psi_i\rangle,
\end{equation} 
where, $E_i$ denote the energy eigenvalue corresponding to the eigenstate $\lvert \psi_i\rangle$ of $\hat{H}$. If $\hat{A}$ is an observable of interest, then the infinite time average of its expectation value is
\begin{eqnarray}\label{sec_two_6}
\bar{A}\equiv\langle\langle \Psi(t)\lvert \hat{A} \lvert\Psi(t)\rangle\rangle_t &\equiv&\lim_{\tau\to\mathcal{1}}~\frac{1}{\tau}~\int_{0}^{\tau} A_t~dt\nonumber\\
&=&\sum_i~\lvert C_i\lvert^2A_{ii},
\end{eqnarray}
where, $A_t$ is the expectation value of the observable at time $t$, given by
\begin{eqnarray}\label{sec_two_7}
A_t&\equiv&\langle\Psi(t)\lvert\hat{A}\lvert \Psi(t)\rangle\nonumber\\
&=&\sum_{ij}~C_i^{\ast}C_j\mbox{e}^{i({E_i-E_j})t}A_{ij}.
\end{eqnarray}
In order to show that $\bar{A}$ is equal to the thermal average, it is assumed that in a chaotic quantum system the matrix elements of $\hat{A}$ take the form\cite{MarkSrednicki_1999}
\begin{equation}\label{sec_two_8}
A_{ij}=\mathcal{A}(E)\delta_{ij}+\mbox{e}^{-S(E)/2}f(E,\omega)R_{ij},
\end{equation}
where, $E\equiv\frac{1}{2}(E_i+E_j)$, $\omega\equiv(E_i-E_j)$, and $S(E)$ is the thermodynamic entropy at energy $E$, given by  
\begin{equation}\label{sec_two_9}
\mbox{e}^{S(E)}\equiv E\sum_i~\delta(E-E_i).
\end{equation}
The functions $f(E,\omega)$ and $\mathcal{A}(E)$ are smooth functions of their arguments. $R_{ij}$ is a numerical factor that varies erratically with $i$ and $j$. Detailed discussion about these functions is beyond the scope of this paper. However, here we focus on few things\cite{MarkSrednicki_1999} that will serve the purpose. First of all, Eq.~(\ref{sec_two_8}) is semiclassical in nature and the factor of $\mbox{e}^{-S(E)/2}$ scales like $\hbar^{(D-1)/2}$. Thus for the validity of this equation, $\hbar$ has to be small, which in practice requires that energy $E$ must be large. Secondly, the general structure described by Eq.~(\ref{sec_two_8}) is preserved under multiplication\cite{MarkSrednicki_1999}, implying the generic character of the equation. As a result, the validity of this equation guarantees the validity of the following expression for the matrix elements of any power of $\hat{A}$,
\begin{equation}\label{sec_two_10}
(A^n)_{ij}=\mathcal{A}_n(E)\delta_{ij}+\mbox{e}^{-S(E)/2}f_n(E,\omega)R^{(n)}_{ij}.
\end{equation}
Thirdly, the function $\mathcal{A}(E)$ can be related to the canonical thermal average of $\hat{A}$ as,
\begin{equation}\label{sec_two_11}
\mathcal{A}(E)=\langle\hat{A}\rangle_{can}+\mathcal{O}(D^{-1})+\mathcal{O}(\mbox{e}^{-S/2}).
\end{equation}
Finally, using Eqs.~(\ref{sec_two_5}-\ref{sec_two_11}) and considering few physical conditions one can show that at thermal equilibrium $\bar{A}=\langle \hat{A}\rangle_{can}$. In the next section, we will show that entanglement between the energy eigenstates of the unperturbed Hamiltonian can facilitate thermalization in an isolated quantum many-body system.

\section{Results}\label{res}
Let the initial Hamiltonian of an isolated quantum many-body system be $\hat{H}_0$ and the dimension of its Hilbert space be $D$. Without loss of generality, we can map the many-body system to a two-body system, $S$ and $R$. Therefore, the initial Hamiltonian can be split as $\hat{H}_0=\hat{H}_S\otimes \mathbf{1}_R+\mathbf{1}_S\otimes\hat{H}_R$. The structure of the eigenstates forming the eigenbasis of $\hat{H}_0$ is as follows:
\vspace{5pt}
\begin{eqnarray}\label{unperturbed eigen}
    \lvert\phi_{lk}\rangle=\lvert\alpha_l\rangle\otimes\lvert\beta_k\rangle,\nonumber\\
    \hat{H}_S\lvert\alpha_l\rangle=a_l\lvert\alpha_l\rangle,~~~~l=1,...,n\nonumber\\
    \hat{H}_R\lvert\beta_k\rangle=b_k\lvert\beta_k\rangle,~~~~k=1,...,m\nonumber\\
    \hat{H}_0\lvert\phi_{lk}\rangle=(a_l+b_k)\lvert\phi_{lk}\rangle,~~~~\forall (l,k)\nonumber\\
    \hat{H}_0\lvert\phi_\nu\rangle=E_\nu\lvert\phi_\nu\rangle,~~~~(l,k)\mapsto \nu,~~ E_\nu=(a_l+b_k).
\end{eqnarray}
In the last equation, we introduce a new index $\nu$ such that it is in a one-to-one correspondence with the original index $(l,k)$. Thus the eigenbasis of the non-interacting Hamiltonian consists of $D$ eigenstates, each being denoted as $\lvert\phi_\nu\rangle$, where $\nu=1,...,nm$ and $nm=D$. Let $E$ be the mean energy of the system. With the hope of making the system obey quantum statistical mechanics a small perturbation $\hat{H}_{int}$ is added to it. Rather than explicitly incorporating these interactions, we opt to represent $\hat{H}_{int}$ as a real symmetric matrix, with its elements randomly sampled from a Gaussian distribution and due to this external perturbation entanglement is established between the energy eigenstates of $S$ and $R$. Before the application of the perturbation, let the total system be in an eigenstate of the noninteracting Hamiltonian $\hat{H}_0$. Instead of raising the initial product state of the system, say with energy $E_\nu$, to an excited state having energy $E_{\nu+1}$, where $E_{\nu+1}>E_\nu$, the external perturbation couples the energy eigenstates of the subsystems. Mathematically speaking $\langle\phi_{\nu}\lvert\hat{H}_{int}\lvert\phi_{\nu}\rangle\approx 0,~~\forall \nu$, which is in agreement with the theory of random matrix\cite{c5bd8f0f-2576-3f83-a184-791e55682183,c9eb8278-b5e2-37bc-a322-8e81785f98ed,Reimann_2015,10.21468/SciPostPhys.15.1.024}. The total Hamiltonian of the system is now $\hat{H}=\hat{H}_0+\hat{H}_{int}$ and the system evolves under this Hamiltonian. The eigenvectors forming the eigenbasis of $\hat{H}$ are the entangled states $\lvert\psi_i\rangle$\cite{Calabrese_2005}. 

The eigenvectors $\{|\psi_i\rangle\}$ of the total Hamiltonian $\hat{H}$ and  $\{|\phi_\nu\rangle\}$ of the unperturbed Hamiltonian $\hat{H}_0$ are related as 
\begin{equation}\label{entangled basis}
    \lvert\psi_i\rangle=\sum_{\nu}~p_{i\nu}\lvert\phi_\nu\rangle,
\end{equation}
which means that $i^{th}$ eigenstate of $\hat{H}$ is a coherent mixture of the eigenstates of $\hat{H}_0$. Thus the transformation of energy eigenstates due to the external perturbation can be viewed as a linear mapping. The elements of the matrix representing the map \textit{i.e.,} the overlaps between unperturbed and perturbed eigenstates are the complex entities $p_{i\nu}\equiv\langle\phi_{\nu}\lvert\psi_{i}\rangle$, square of which denote the probability of $\nu$-th eigenstate of unperturbed Hamiltonian in the $i$-th eigenstate of the total Hamiltonian. The overlaps $p_{i\nu}$s are chosen from zero-mean Gaussian distribution and are statistically independent of each other. Therefore, normalization condition requires $\sum_{\nu}~\lvert p_{i\nu}\lvert^2=1$. Let us denote the transformation matrix as $P$, and assume it to be a random unitary matrix. The matrix corresponding to the inverse map will therefore be $P^{\dagger}$, because for unitary matrix $P^{\dagger}=P^{-1}$. Since $P$ is a joint unitary operator acting on the Hilbert space $\mathcal{H}_S\otimes\mathcal{H}_R$, where $\mathcal{H}_{S(R)}$ is the Hilbert space of $S(R)$, it can transform a product state either into an entangled state or another product state. However, given that $P$ is randomly chosen and the set of product states has measure zero in $\mathcal{H}_S\otimes\mathcal{H}_R$, the probability of $|\psi_i\rangle$ being a product state is exceedingly small. In other words, $|\psi_i\rangle$ is almost always an entangled state\cite{Calabrese_2005}.
In the following subsections, we show the revival of the results of quantum statistical mechanics.

\subsection{Microcanonical average}
As we want to show that due to entanglement the isolated system will move towards thermal equilibrium, we need to look at the variation in the expectation value of an observable (self-adjoint operator) $\hat{A}$ in an energy eigenstate,
\begin{equation}\label{micro1}
	\langle\psi_i\lvert \hat{A}\lvert\psi_i\rangle=\sum_{\nu\mu}~p_{i\nu}^{\ast}p_{i\mu}\langle\phi_{\nu}\lvert \hat{A}\lvert\phi_{\mu}\rangle.
\end{equation}
In the limit of large $D$ there will be numerous energy levels in the window $\Delta E$ having different eigenstates. Therefore, the expectation value will vary from state to state. In such a case \textit{variance} is a good measure to find out how the expectation value in an eigenstate differs from a mean value. For this, we first calculate the average value of $\langle\psi_i\lvert \hat{A}\lvert\psi_i\rangle$. We find (see Appendix \ref{appenA.1})
\begin{eqnarray}\label{micro2}
	\langle\langle\psi_i\lvert \hat{A}\lvert\psi_i\rangle\rangle_{rand}=\sum_{\nu}~\frac{1}{D_{E'_{i}}}\langle\phi_{\nu}\lvert \hat{A}\lvert\phi_{\nu}\rangle.
\end{eqnarray} 
Here, $D_{E'_{i}}$ is the Hilbert space dimension of the energy shell $[E'_{i}, E'_{i}\pm \Delta E'_{i}]$, {\it i.e.,} $D_{E'_{i}}$ represents the number of energy eigenstates contained in the mentioned energy shell.

Considering the macroscopic energy of the system as $E$, we can denote $I_{micro}\coloneqq[E,E\pm\Delta E]$ as the usual microcanonical energy shell, whose spread $\Delta E$ is macroscopically small, \textit{i.e.,} beyond the experimental resolution limit but microscopically large. Microscopically large means much larger than the typical energy level spacing $(E_{j+1}-E_{j})$. The number of energy eigenvalues $E_{j}$ contained in $I_{micro}$ is the degrees of freedom $D$ and is typically very large. The corresponding microcanonical ensemble is given by 
\begin{equation}\label{micro3}
	\rho_{micro}\coloneqq \dfrac{1}{D}\sum_{j}\lvert\phi_j\rangle\langle\phi_j\lvert
\end{equation}
where, the sum $\sum_{j}$ is over all $j$ such that the energy eigenvalue $E_{j}$ lies within $I_{micro}$.

Thus the expectation value of $\hat{A}$ in the microcanonical ensemble takes the form
\begin{eqnarray}\label{micro4}
	\langle \hat{A}\rangle_{micro}(E)&=&\mbox{Tr}[\rho_{micro}\hat{A}] \nonumber\\
	&=&\sum_{j}\frac{1}{D}\mbox{Tr}[\lvert\phi_j\rangle\langle\phi_j\lvert \hat {A}]\nonumber\\
	&=&\sum_j~\frac{1}{D}\langle\phi_j\lvert \hat{A}\lvert\phi_j\rangle.
\end{eqnarray}

Hence, we find that the average of the expectation value of the observable $\hat{A}$ in an eigenstate of the global system equals the thermal (microcanonical) average $\langle \hat{A}\rangle_{micro}(E_{i}')$ of $\hat{A}$ at the mean energy $E'_{i}$, \textit{i.e.,}
\begin{equation}\label{micro5}
	\langle\langle\psi_i\lvert \hat{A}\lvert\psi_i\rangle\rangle_{rand}=\langle \hat{A}\rangle_{micro}(E'_{i}).
\end{equation}
This postulates that the energy eigenstates of an isolated quantum many-body system are thermal, \textit{i.e.,} each of them yields practically the same expectation values as the microcanonical ensemble.

The variance is found to be (Appendix \ref{appenA.2})
\begin{eqnarray}\label{micro6}
	\Delta\hat{A}^2&=&\langle\langle\psi_i\lvert \hat{A}\lvert\psi_i\rangle^2\rangle_{rand}-\langle\langle\psi_i\lvert \hat{A}\lvert\psi_i\rangle\rangle^2_{rand}\nonumber\\
	&\leq& \frac{2}{D_{E'_{i}}}\langle \hat{A}^2\rangle_{micro}(E'_{i}).
\end{eqnarray}
In the limit of large $D_{E'_{i}}$, the fluctuation in the expectation value of the observable $\hat{A}$ in an energy eigenstate of interacting Hamiltonian becomes negligible, which is clear from the above equation.

Now, let us consider an arbitrary initial state of the perturbed system,
\begin{equation}\label{micro7}
	\lvert\Psi\rangle=\sum_{i}~C_{i}\lvert\psi_{i}\rangle,
\end{equation}
where, $\sum_i~ \lvert C_i\lvert^2=1$. To find out the time average of the expectation value, the arbitrary initial state of such perturbed quantum system at any instant of time $t$ is
\begin{equation}\label{micro8}
	\lvert \Psi(t)\rangle=\sum_{i}~C_i\mbox{e}^{-\mbox{i}E'_i t}\lvert \psi_i\rangle,
\end{equation} 
where, $E'_{i}$ denotes the energy eigenvalue concomitant with the eigenstate $\lvert \psi_i\rangle$ of $\hat{H}$. In the interacting basis, the ensemble-averaged time average of the expectation of $\hat{A}$ can be shown to be (see Appendix \ref{appenA.3})
\begin{eqnarray}\label{micro9}
	\langle\langle\langle\Psi(t)\lvert\hat{A}\lvert\Psi(t)\rangle\rangle_t\rangle_{rand}&\equiv&\lim_{T\to\infty}\frac{1}{T}\int_0^T\langle\langle\Psi(t)\lvert\hat{A}\lvert\Psi(t)\rangle\rangle_{rand}~dt\nonumber\\
	&=&\sum_i~\lvert C_{i}\lvert^2\langle\langle\psi_i\lvert\hat{A}\lvert\psi_i\rangle\rangle_{rand}.
\end{eqnarray}
Now invoking Eqs. (\ref{micro2},\ref{micro4} and \ref{micro5}), we finally get, in the limit of a large number of degrees of freedom,
\begin{align}\label{micro10}
	\langle\langle\langle\Psi(t)\lvert\hat{A}\lvert\Psi(t)\rangle\rangle_{t}\rangle_{rand} &=\sum_{i}~\lvert C_{i}\lvert^2\langle\hat{A}\rangle_{micro}(E'_{i})\nonumber\\
    &=\langle\hat{A}\rangle_{micro}(E)\nonumber\\
    &\equiv \dfrac{1}{D_{E}}\sum_{\substack{\alpha\\ \lvert E-E_{\alpha}\lvert<\Delta E}}\langle\phi_{\alpha}\lvert\hat{A}\lvert\phi_{\alpha}\rangle,
\end{align}
where $\Delta E$ is the spread of the mean energy $E$ of the initial state, and $D_{E}$ is the number of energy eigenstates with energies in the energy shell $[E,E\pm\Delta E]$.

Therefore, the system has finally achieved \textit{ergodicity} due to entanglement. 

\subsection{Canonical average}
Now we consider an arbitrary observable $\hat{M}$, a self-adjoint operator, of the subsystem $S$ and find out the expectation value of this observable. It is assumed that the dimension of $S$ is much smaller than that of $R$. Let us express an eigenstate of the total Hamiltonian $\hat{H}$ as
\begin{equation}\label{cano1}
\lvert\psi\rangle=\sum_{l=1}^n\sum_{k=1}^m~p_{lk}\lvert\alpha_l\rangle\otimes\lvert\beta_k\rangle.
\end{equation} 
Then the expectation value is
\begin{eqnarray}\label{cano2}
\langle\psi\lvert\hat{M}\otimes\mathbf{1}_R\lvert\psi\rangle=\sum_{l,l^{\prime}=1}^n\sum_{k=1}^m~p_{lk}
^{\ast}p_{l^{\prime}k}M_{ll^{\prime}}.
\end{eqnarray}
From quantum mechanical principles, we have,
\begin{equation}\label{cano3}
\langle\psi\lvert\hat{M}\otimes\mathbf{1}_R\lvert\psi\rangle=\mbox{Tr}(\hat{M}\rho_S),
\end{equation}
where, $\rho_S$ is the state of the subsystem $S$ after performing partial trace on the eigenstate $\lvert\psi\rangle$, \textit{i.e.,} 
$\rho_S\equiv\mbox{Tr}_R(\lvert\psi\rangle\langle\psi\lvert)$. As every eigenstate of $\hat{H}$ is entangled, the state of subsystem $S$ is mixed and diagonal in its eigenbasis. The diagonal elements are the terms $\sum_{k=1}^{m}\lvert p_{lk}\lvert^2$, which are generally functions of energy\cite{PhysRevLett.80.1373}. However, we do not express the elements $\rho_{ll}$ explicitly in terms of energy. Rather, we use the result of \cite{popescu2006entanglement}, where it has been shown that using Levy's lemma\cite{milman1986asymptotic} one can prove that $\rho_{ll}\equiv\frac{\mbox{exp}(-\beta
E_l)}{\mathcal{Z}}$ for the condition $m\gg n$ (the dimension of the subsystem $R$ is much larger than that of $S$), where $\mathcal{Z}=\sum_l~\mbox{exp}(-\beta E_l)$ is the \textit{partition function}. The Gibb's form\cite{gibbs1902elementary} of density matrix depends on the nature of coupling between $S$ and $R$; for weak coupling $\rho_S\equiv\mbox{exp}(-\beta\hat{H}_S)/\mathcal{Z}$, whereas, for strong coupling $\rho_S\equiv\mbox{exp}(-\beta\hat{H}^{\ast}_S)/\mathcal{Z}^{\ast}$, where $\hat{H}^{\ast}_S$ is the Hamiltonian of \textit{mean force} and $\mathcal{Z}^{\ast}=\mbox{Tr}_{S}\{\mbox{exp}(-\beta\hat{H}^{\ast}_S)\}$\cite{PhysRevLett.102.210401,PhysRevE.84.031110}. Thus, in accordance with the guidelines outlined in \cite{PhysRevLett.80.1373}, it follows that for any generic state of $S$,
\begin{eqnarray}\label{cano4}
\langle\psi\lvert\hat{M}\otimes\mathbf{1}_R\lvert\psi\rangle&=&\sum_{l,l^{\prime}=1}^n\sum_{k=1}^m~p_{lk}
^{\ast}p_{l^{\prime}k}M_{ll^{\prime}}\nonumber\\
&\simeq&\dfrac{\sum_{l}\sum_{k}~\lvert p_{lk}\lvert^{2}M_{ll}}{\sum_{l}\sum_{k}~\lvert p_{lk}\lvert^{2}}\nonumber\\
&=&\dfrac{\sum_l~\rho_{ll}M_{ll}}{\sum_l~\rho_{ll}}\nonumber\\
&=&\dfrac{\mbox{Tr}(\hat{M}\rho_S)}{\mbox{Tr}(\rho_S)}\nonumber\\
&\equiv&\langle M\rangle_{can}.
\end{eqnarray}
Therefore, we find that due to entanglement, not only does the global system approach thermodynamic equilibrium, but the subsystem $S$ also undergoes thermalization. The initial state of the subsystem loses its purity and the expectation value of any observable becomes equivalent to the canonical average.

\section{Conclusion}\label{con} 
We have revisited the problem of thermalization of a generic isolated quantum many-body system and shown that the establishment of entanglement in the system leads to thermalization, which is in confirmation with experimental evidence\cite{doi:10.1126/science.aaf6725}. According to ETH \textit{ergodicity} arises due to the coupling of neighboring energy levels in the window $\Delta E$ or coherent mixing of eigenstates of unperturbed Hamiltonian $\hat{H}_0$. 
We have looked at this mixing of eigenstates from a different perspective. In a quantum many-body system, when there is no interaction between the subsystems, an eigenstate of $\hat{H}_0$ is basically a product state, where the components of the product are the eigenstates of the subsystems. Now, when these product states mix coherently, the resulting state is an entangled state and an eigenstate of the interacting Hamiltonian $\hat{H}$. We have considered one such entangled eigenstate and analyzed the expectation value of an observable $\hat{A}$ in that state. We have found that entanglement not only gives rise to thermalization, it thermalizes the system according to \textit{eigenstate thermalization hypothesis}. We have shown that in the limit of a large number of degrees of freedom, the fluctuation in the expectation value in an eigenstate becomes negligibly small and the expectation value is equivalent to the microcanonical average.
 Our result shows that the time average of the expectation value is also equivalent to the microcanonical average. Basically, time averaging plays an auxiliary role; thermalization happens at the level of individual eigenstates and the expectation value of an observable is equivalent to its equilibrium value in any eigenstate. To check the equivalence between the expectation value of an observable and canonical average we consider an observable of a subsystem and find its expectation value. As the global pure system gets entangled, the initial pure state of any subsystem becomes mixed, which becomes clear when we take partial trace to find the density matrix $\rho_S$. Experimentally, the same fact has been observed, entanglement starts to grow after a quench is applied to a closed and pure many-body quantum system and destroys the purity of the subsystems. The subsystems become mixed and the second-order R\'enyi entropy can quantify their mixedness as $S(A)=-\mbox{log}[\mbox{Tr}(\rho)_A]$\cite{doi:10.1126/science.aaf6725}. Instead of determining the elements of $\rho_S$ explicitly as a function of energy, we have used the known result: if weakly coupled subsystems $S$ and $R$ are entangled, and the dimension of $R$ is much larger than that of $S$ then $\rho_S\equiv\rho_{can}$. Using this fact we have found that the expectation value of observable $\hat{M}$ is equivalent to canonical average. Previous theoretical works\cite{PhysRevLett.125.070605} have also noticed the relation between entanglement and thermalization of an isolated quantum system by finding that the function $f(E,\omega)$ in Eq.(\ref{sec_two_8}) carries multipartite entanglement structure of the energy eigenstates. Though our work represents implicit relation between entanglement and thermalization, we hope that this will motivate the further investigation of the role of entanglement energetics in the thermalization of an isolated quantum many-body system, thereby paving the path for better understanding of quantum thermodynamics\cite{deffner2019quantum,PhysRevE.85.061126,e15062100,PhysRevE.65.055102,doi:10.1073/pnas.1411728112,masanes2017general,lewis2019unifying,e20060423}. 

\section*{Acknowledgements}
TS and PG acknowledge Prof. Sibasish Ghosh for the insightful discussions. PG acknowledges his academic visit to IMSc, Chennai, where fruitful discussions regarding this work took place. PB and AB would like to thank Bose Institute for providing support to carry out this work. PD would like to thank the Council of Scientific and Industrial Research, Govt. of India, for the financial support through the Research Associateship award (EMR-1/2018/0527).

\appendix
\section{Appendix} 
\label{appenA}
\subsection{Microcanonical average}\label{appenA.1}
Using the notation $\langle \cdot\cdot\cdot \rangle_{rand}$ to denote the average over the random matrix ensemble, we employ
\begin{equation}\label{random gaussian condn1}
\Lambda(i,\nu)\equiv\langle\lvert p_{i\nu}\lvert^{2}\rangle_{rand}=\Lambda(i-\nu),~~~~\sum_{n}\Lambda(n)=1,~~~~max_{n}\Lambda(n)=\mathcal{O}(e^{-N}),
\end{equation}
where the $\Lambda(n)$ are real (but not necessarily even) function of $n$. $\Lambda(n)$ is monotonically decreasing for $n\geq 0$ and monotonically increasing for $n\leq 0$, hence exhibiting a global maximum at $n=0$. This holds for both the Gaussian orthogonal or unitary ensemble (and even beyond strict Gaussianity)\cite{Reimann_2015,10.21468/SciPostPhys.15.1.024}. The important point is that the overlap between eigenvectors of $\hat{H}$ and $\hat{H_{0}}$ is exponentially small in the particle number $N$ and for our analysis, we will set for notational simplicity $max_{n} \Lambda(n)=(1/D)$ with $D$ the Hilbert space dimension of the pertinent energy shell\cite{10.21468/SciPostPhys.15.1.024}.
 
The $p_{i\nu}$s are independent zero-mean Gaussian random numbers obeying 
\begin{equation}\label{random gaussian condn2}
\langle p_{i\mu} \rangle_{rand}=0,~~~~\langle p_{i\mu}p_{j\nu} \rangle_{rand}=0=\langle p^{*}_{i\mu}p^{*}_{j\nu} \rangle_{rand},~~~\langle p_{i\mu}p^{*}_{j\nu} \rangle_{rand}=\delta_{ij}\delta_{\mu\nu}~\Lambda(i-\mu).
\end{equation}
Again, the ensemble average of $2n$ random variables can be calculated using \textit{Isserlis theorem}, which sums over pairing where each pairing is a product of $n$ pairs, \textit{e.g.,} 
\begin{align}\label{random gaussian condn3}
\langle p_{i\nu}p^{*}_{i\mu}p_{il}p^{*}_{im} \rangle_{rand}&=\langle p_{i\nu}p^{*}_{i\mu} \rangle_{rand}\langle p_{il}p^{*}_{im} \rangle_{rand}+\langle p_{i\nu}p^{*}_{im} \rangle_{rand}\langle p_{il}p^{*}_{i\mu} \rangle_{rand}\nonumber\\
&=\delta_{\nu\mu}\delta_{lm}\Lambda(i-\nu)\Lambda(i-l)+\delta_{\nu m}\delta_{l\mu}\Lambda(i-\nu)\Lambda(i-l)\nonumber\\
&= \dfrac{1}{D_{E'_{i}}^{2}}\big\{\delta_{\nu\mu}\delta_{lm}+\delta_{\nu m}\delta_{l\mu}\big\},
\end{align}
where, $D_{E'_{i}}$ is the Hilbert space dimension of the energy shell $[E'_{i},E'_{i}\pm\Delta E'_{i}]$.

Now, the ensemble average of the expectation value of $\hat{A}$ is
\begin{align}
	\langle\langle\psi_i\lvert \hat{A}\lvert\psi_i\rangle\rangle_{rand}&=\sum_{\nu\mu}~\langle p^{*}_{i\nu}p_{i\mu}\rangle_{rand}~\langle\phi_{\nu}\lvert \hat{A}\lvert\phi_{\mu}\rangle\nonumber\\
	&= \sum_{\nu\mu}~\delta_{\nu\mu}~\Lambda(i-\nu)~\langle\phi_{\nu}\lvert \hat{A}\lvert\phi_{\mu}\rangle\nonumber\\
	&= \frac{1}{D_{E'_{i}}}\sum_{\nu\mu}~\delta_{\nu\mu}~\langle\phi_{\nu}\lvert \hat{A}\lvert\phi_{\mu}\rangle\nonumber\\
	&= \sum_{\nu}~\frac{1}{D_{E'_{i}}}~\langle\phi_{\nu}\lvert \hat{A}\lvert\phi_{\nu}\rangle.
\end{align}
The right-hand side of the above equation is equivalent to the microcanonical average of the observable $\hat A$ at the mean energy $E'_{i}$, \textit{i.e.,} $\langle\hat{A}\rangle_{micro}(E'_{i})$, as depicted in Eq.~(\ref{micro4}). This result holds for an individual initial state; in contrast to the classical theory, no averaging over an ensemble of initial states is needed.

\subsection{Variance in expectation value}\label{appenA.2}
The variance of expectation value of $\hat{A}$ is
\begin{equation}
	\Delta\hat{A}^2=\langle\langle\psi_i\lvert \hat{A}\lvert\psi_i\rangle^2\rangle_{rand}-\langle\langle\psi_i\lvert \hat{A}\lvert\psi_i\rangle\rangle^2_{rand}.\nonumber
\end{equation}
The first term on the right-hand side can be evaluated by substituting Eq.~(\ref{entangled basis}) for $\lvert\psi_{i}\rangle$ and averaging as before. We have,
\begin{align}
\langle\langle\psi_{i}\lvert\hat{A}\lvert\psi_{i}\rangle^{2}\rangle_{rand}&=\sum_{\nu\mu lm}\langle p^{*}_{i\nu}p_{i\mu}p^{*}_{il}p_{im}\rangle_{rand}\langle\phi_{\nu}\lvert\hat{A}\lvert\phi_{\mu}\rangle\langle\phi_{l}\lvert\hat{A}\lvert\phi_{m}\rangle\nonumber\\
&=\sum_{\nu\mu lm}\big\{\langle p^{*}_{i\nu}p_{i\mu} \rangle_{rand}\langle p^{*}_{il}p_{im} \rangle_{rand}+\langle p^{*}_{i\nu}p_{im} \rangle_{rand}\langle p^{*}_{il}p_{i\mu} \rangle_{rand}\big\}\langle\phi_{\nu}\lvert\hat{A}\lvert\phi_{\mu}\rangle\langle\phi_{l}\lvert\hat{A}\lvert\phi_{m}\rangle\nonumber\\
&=\sum_{\nu\mu lm}\dfrac{1}{D_{E'_{i}}^2}\big\{\delta_{\nu\mu}\delta_{lm}+\delta_{\nu m}\delta_{l\mu}\big\}\langle\phi_{\nu}\lvert\hat{A}\lvert\phi_{\mu}\rangle\langle\phi_{l}\lvert\hat{A}\lvert\phi_{m}\rangle\nonumber\\
&=\dfrac{1}{D_{E'_{i}}^2}\bigg\{\sum_{\nu l}\langle\phi_{\nu}\lvert\hat{A}\lvert\phi_{\nu}\rangle\langle\phi_{l}\lvert\hat{A}\lvert\phi_{l}\rangle+\sum_{\nu l}\langle\phi_{\nu}\lvert\hat{A}\lvert\phi_{l}\rangle\langle\phi_{l}\lvert\hat{A}\lvert\phi_{\nu}\rangle\bigg\}\nonumber\\
&=\dfrac{1}{D_{E'_{i}}^2}\bigg\{\Big[Tr\big(\hat{A}\big)\Big]^{2}+\sum_{\nu}\langle\phi_{\nu}\lvert\hat{A}^{2}\lvert\phi_{\nu}\rangle\bigg\}.
\end{align}
Therefore, we finally get,
\begin{align}
\Delta\hat{A}^2&=\dfrac{1}{D_{E'_{i}}^2}\Big[Tr\big(\hat{A}\big)\Big]^{2}+\dfrac{1}{D_{E'_{i}}^2}\sum_{\nu}\langle\phi_{\nu}\lvert\hat{A}^{2}\lvert\phi_{\nu}\rangle-\dfrac{1}{D_{E'_{i}}^2}\Big[Tr\big(\hat{A}\big)\Big]^{2}\nonumber\\
&=\dfrac{1}{D_{E'_{i}}^2}\sum_{\nu}\langle\phi_{\nu}\lvert\hat{A}^{2}\lvert\phi_{\nu}\rangle\nonumber\\
&=\dfrac{1}{D_{E'_{i}}}\langle\hat{A}^2\rangle_{micro}(E'_{i})\nonumber\\
&\leq \dfrac{2}{D_{E'_{i}}}\langle\hat{A}^2\rangle_{micro}(E'_{i}).
\end{align}

\subsection{Time average}\label{appenA.3}
Again, the ensemble-averaged time average of the expectation value is
\begin{align}
		\langle\langle\langle\Psi(t)\lvert\hat{A}\lvert\Psi(t)\rangle\rangle_{t}\rangle_{rand}&\equiv\lim_{T\to\infty}\frac{1}{T}\int_0^T\langle\langle\Psi(t)\lvert\hat{A}\lvert\Psi(t)\rangle\rangle_{rand}~dt\nonumber\\
	&=\lim_{T\to\infty}\frac{1}{T}\int_0^T\sum_{ik}C_{i}^{*}C_{k}~\mbox{e}^{\mbox{i}\big(E'_{i}-E'_{k}\big)t}~\langle\langle\psi_{i}\lvert\hat{A}\lvert\psi_{k}\rangle\rangle_{rand}~dt\nonumber\\
	&= \lim_{T\to\infty}\frac{1}{T}\int_0^T\Big\{\sum_{i}\lvert C_{i}\lvert^{2}\langle\langle\psi_{i}\lvert\hat{A}\lvert\psi_{i}\rangle\rangle_{rand}+\sum_{i\neq k}C_{i}^{*}C_{k}~\mbox{e}^{\mbox{i}\big(E'_{i}-E'_{k}\big)t}\langle\langle\psi_{i}\lvert\hat{A}\lvert\psi_{k}\rangle\rangle_{rand}\Big\}dt\nonumber\\
	&= \sum_{i}~\lvert C_{i}\lvert^{2}~\langle\langle\psi_{i}\lvert\hat{A}\lvert\psi_{i}\rangle\rangle_{rand}\nonumber\\
	&= \sum_{i}~\lvert C_{i}\lvert^{2}\langle\hat{A}\rangle_{micro}(E'_{i})
\end{align}
In the above equation, the second term of the third line vanishes as the phases cancel out each other in the limit of a large number of degrees of freedom.

\bibliographystyle{JHEP}
\bibliography{biblio.bib}

\end{document}